\documentclass[12pt]{article}
\usepackage{cite} 
\usepackage{amsfonts,latexsym,amssymb}
\usepackage{epsfig}
\begin{document}
\newcommand{\de}{\delta}\newcommand{\ga}{\gamma}
\newcommand{\e}{\epsilon} \newcommand{\ot}{\otimes}
\newcommand{\be}{\begin{equation}} \newcommand{\ee}{\end{equation}}
\newcommand{\ba}{\begin{array}} \newcommand{\ea}{\end{array}}
\newcommand{\beq}{\begin{equation}}\newcommand{\eeq}{\end{equation}}
\newcommand{\tmod}{{\cal T}}\newcommand{\amod}{{\cal A}}
\newcommand{\bemod}{{\cal B}}\newcommand{\cmod}{{\cal C}}
\newcommand{\dmod}{{\cal D}}\newcommand{\hmod}{{\cal H}}
\newcommand{\s}{\scriptstyle}\newcommand{\tr}{{\rm tr}}
\newcommand{\einsop}{{\bf 1}}
\def\R{\overline{R}} \def\doa{\downarrow}
\def\dag{\dagger}
\def\ve{\epsilon}
\def\si{\sigma}
\def\ga{\gamma}
\def\nn{\nonumber}
\def\le{\langle}
\def\re{\rangle}
\def\lt{\left}
\def\rt{\right}
\def\dwn{\downarrow}   
\def\up{\uparrow}
\def\dag{\dagger}
\def\bea{\begin{eqnarray}}
\def\eea{\end{eqnarray}}
\def\p{\tilde{p}}
\def\q{\tilde{q}}
\def\H{\overline{H}}
\newcommand{\reff}[1]{eq.~(\ref{#1})}

\title{Classical and quantum dynamics of a model for atomic-molecular 
Bose--Einstein condensates}

\author{ G. Santos$^1$, A.Tonel$^1$, A. Foerster$^1$ and J. Links$^2$
\vspace{1.0cm}\\
$^{1}$ Instituto de F\'{\i}sica da UFRGS \\ 
Av. Bento Gon\c{c}alves 9500, Porto Alegre, RS - Brazil
\vspace{0.5cm}\\
$^{2}$ Centre for Mathematical Physics, School of Physical Sciences \\ 
The University of Queensland, Queensland, 4072, Australia}

\maketitle

\begin{abstract}
We study a model for a two-mode 
atomic-molecular Bose--Einstein condensate. Starting with a classical analysis we determine the 
phase space fixed 
points of the system. It is found that bifurcations of the fixed points naturally separate the coupling parameter space 
into four regions. The different regions give rise to qualitatively different dynamics. We then show that this classification holds true for the quantum dynamics. 
% Using  
% direct numerical diagonalisation of the Hamiltonian we compute the time evolution of the 
% pectation value of the relative atom number for different values of the couplings, showing that 
% e system exhibits rich and complex behaviours.

\end{abstract}


\vfil\eject
%%%%%%%%%%%%%%%%

\section{Introduction}

Since the early experimental realisations of Bose--Einstein condensates (BECs) using alkali atoms \cite{cw,ak}, 
a significant effort has been made to produce a stable BEC in a gas of molecules \cite{z}. A molecular 
condensate could lead to a host of new scientific investigations that includes the quantum gas 
with anisotropic dipolar interactions \cite{[2a]}, 
% the test of fundamental symmetries such as the search for a permanent eletric 
% dipole moment \cite{dipole}, 
the study of rotational and vibrational energy transfer processes \cite{rv} and coherent chemistry 
where the reactants and products are in a coherent quantum superposition of states \cite{heinzen}, 
among others.
In recent years the creation of a molecular BEC from an atomic BEC 
has been achieved  by different techniques such as photoassociation \cite{[5a]}, two-photon Raman 
transition \cite{wynar} and Feshbach resonance \cite{[7]}.

{}From a theoretical point of view, molecular BECs may be studied using  the Gross-Pitaevski (GP) equations
and mean-field theory (MFT) (e.g. see \cite{vardi,caok}).  
The GP-MFT approach reduces the full multi-body problem
into a set of coupled nonlinear Schr\"odinger equations, which are then solved 
numerically to obtain the Josephson-type
dynamics of the coupled atomic and molecular fields. 
An approximation can be made to reduce the complex multi-body problem 
into a two-mode problem. 
An analysis of this two-mode Hamiltonian was carried out in 
\cite{vardi}, where
it was established that  
the quantum solutions
break away from the MFT predictions in the vicinity of the dynamically unstable molecular 
mode due to strong quantum
fluctuations. 
It has been   
shown that the two-mode Hamiltonian is an exactly solvable model in the framework of the algebraic Bethe
ansatz method \cite{lzmg} 
and an analysis using these results was given in \cite{zlm}.  
However in most of the above investigations, the 
atom-atom, atom-molecule and molecule-molecule $S$-wave scattering interactions were not taken into account.

In the present work we focus on a more general Hamiltonian
which takes into account the $S$-wave scattering interactions.
By means of a classical analysis we first
obtain the fixed 
points of the system and find that the space of coupling parameters 
divides into four distinct regions which are determined by 
fixed point bifurcations. By contrast, 
only three such regions exist when the $S$-wave scattering interactions are neglected.
The results allow us to qualitatively predict the dynamical behaviour of the system in terms of whether
the evolution is localised or delocalised.
Using exact diagonalisation of the Hamiltonian, 
we then see that the quantum dynamics within each region has a similar character.

The paper is organised as follows: In section 2 we present the Hamiltonian and 
in section 3 a classical analysis of the model is performed.
In section 4 we investigate the quantum dynamics through 
the time evolution of the expectation value of the relative 
atom number. Section 5 is reserved for a discussion of the results.

\section{The model}

Let us consider the following general Hamiltonian, based on the two-mode approximation, describing 
the coupling between atomic and diatomic-molecular Bose-Einstein 
condensates 
\begin{equation}
H=U_aN_a^2 + U_bN_b^2 +U_{ab}N_aN_b + \mu_aN_a +\mu_bN_b + \Omega(a^{\dag}a^{\dag}b +b^{\dag}aa).
\label{ham}
\end{equation}
Above, $a^{\dagger}$ is the creation operator for an atomic mode while $b^{\dagger}$ 
creates a molecular mode. The Hamiltonian commutes
with the total atom number $N=N_a+2N_b$,
where $N_a=a^{\dagger}a$ and $N_b=b^{\dagger}b$.
Notice that the change of variable $\Omega\rightarrow -\Omega$ is equivalent to the unitary transformation
\begin{eqnarray}
 b\rightarrow - b. \label{trans}
\end{eqnarray} 

The parameters $U_{j}$ describe S-wave scattering, taking into 
account the atom-atom ($U_{a}$), atom-molecule ($U_{ab}$) and molecule-molecule ($U_{b}$) interactions. 
The parameters $\mu_i$ are external potentials 
and $\Omega$ is the amplitude for interconversion of atoms and molecules. 
In the limit $U_{a}=U_{ab}=U_{b}=0$, (\ref{ham}) has been studied using a
variety of methods \cite{vardi,lzmg,zlm,hmm03}. However in the experimental context, the $S$-wave 
scattering interactions play a significant role. It will be seen below that for the  
general model (\ref{ham}) the inclusion of these scattering terms has a non-trivial consequence.
We mention that generally the values for 
$U_b$ and $U_{ab}$ are unknown \cite{wynar,heinzen}, although some estimates
exist in the case of $^{85}Rb$ \cite{caok}.

We finally note that the Hamiltonian (\ref{ham}) is a natural generalisation of the two-site Bose-Hubbard model
\begin{equation}
H=U(N_1-N_2)^2 +\mu(N_1-N_2)+ \Omega(a^{\dag}_1a_2 +a_2^{\dag}a_1)
\label{bh}
\end{equation}
which has been extensively studied as a model for quantum tunneling between two 
single-mode Bose--Einstein condensates \cite{hmm03,mcww,rsk,leggett,ks,our,ours}. 
Our analysis will show that 
despite apparent similarities between the Hamiltonians (\ref{ham}) and (\ref{bh}), they do display some very different properties.
This aspect will be discussed in Section \ref{discussion}.

\section{The classical analysis}

Let $N_j,\,\theta_j,\,j=a,\,b$ be
quantum variables satisfying the canonical relations 
$$[\theta_a,\,\theta_b]=[N_a,\,N_b]=0,~~~~~[N_j,\,\theta_k]=i\delta_{jk}I.$$  
Using the fact that 
$$\exp(i\theta_j)N_j=(N_j+1)\exp(i\theta_j) $$ 
we make a change of variables from the operators $j,\,j^\dagger,\,j=a,\,b$ via
$$j=\exp(i\theta_j)\sqrt{N_j},
~~~j^\dagger=\sqrt{N_j}\exp(-i\theta_j) $$ 
such that the Heisenberg canonical commutation relations are preserved.  
We make a further change of variables
$$ z=\frac{1}{N}(N_a-2N_b),$$
$$ N=N_a+2N_b, $$
$$\theta=\frac{N}{4}(2\theta_a-\theta_b),$$ 
such that $z$ and $\theta$ are canonically conjugate variables; i.e.
$$[z,\,\theta]=iI. $$   
In the limit of large $N$ we can now approximate the (rescaled) Hamiltonian by 
\bea
H=\lambda z^2 +2 \alpha z +\beta   
+\sqrt{2(1-z)}(1+z) \cos\left(\frac{4\theta}{N}\right)
\label{ham2}
\eea
with
\begin{eqnarray*} \lambda &=& \frac{\sqrt{2N}}{\Omega}\left(\frac{U_{a}}{2}
-\frac{U_{ab}}{4}+\frac{U_{b}}{8}
\right)  \\
\alpha &=&\frac{\sqrt{2N}}{\Omega}\left(\frac{U_{a}}{2}
-\frac{U_{b}}{8} + \frac{\mu_a}{2N}-\frac{\mu_b}{4N}\right)   \\
\beta &=& \frac{\sqrt{2N}}{\Omega}\left(\frac{U_{a}}{2}
+\frac{U_{ab}}{4}+\frac{U_{b}}{8}+\frac{\mu_a}{N}+\frac{\mu_b}{2N}
\right)   
\end{eqnarray*} 
where, since $N$ is conserved, we treat it as a constant. 
We note that the unitary transformation (\ref{trans})
is equivalent to $\theta \rightarrow \theta +{N\pi}/{4}$. 
Also, since the Hamiltonian (\ref{ham}) is 
time-reversal invariant, we will hereafter restrict our analysis to the case $\lambda \geq 0$.

\begin{figure}[ht]
\begin{center}
\epsfig{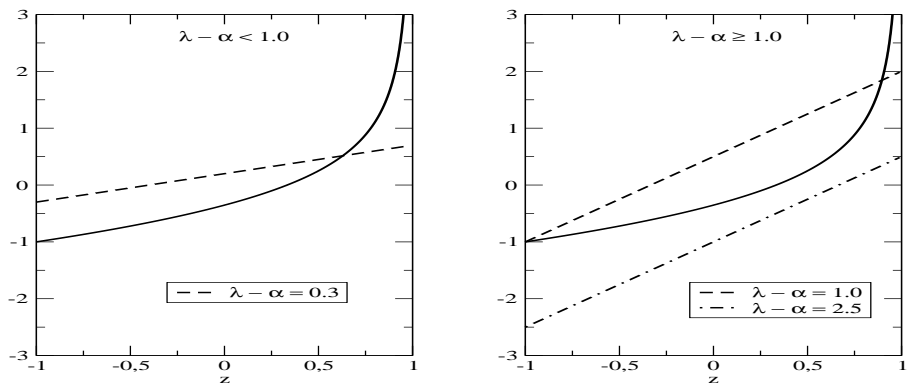}
\caption{ Graphical solution of equation (\ref{sol2}). 
The crossing between the straight line (left hand side of eq.(\ref{sol2})) and the curve
(right hand side of eq.(\ref{sol2})) for different $\lambda - \alpha$ values represents the solution(s)
for each case. There is just one solution on the left  ($\lambda - \alpha < 1$), 
while there are either two solutions or no solution on the right ($\lambda - \alpha \geq 1$). }
\label{fig2}
\end{center}
\end{figure}

We now regard (\ref{ham2}) as a classical Hamiltonian and 
investigate the fixed points of the system. The first step is to 
find Hamilton's equations of motion which yields
\begin{eqnarray*} 
\frac{dz}{dt}=\frac{\partial H}{\partial \theta}&=&-\frac{4}{N}\sqrt{2(1-z)}  
(1+z)  \sin\left(\frac{4\theta}{N}\right),  \label{de1} \\
-\frac{d\theta}{dt}=\frac{\partial H}{\partial z} &=&2\lambda z +2\alpha 
+\frac{1-3z}{\sqrt{2(1-z)}} \cos\left(\frac{4\theta}{N}\right).
\label{de2}  
\end{eqnarray*}
The fixed points of the system are determined by the condition
\begin{equation}
\frac{\partial H}{\partial \theta}=\frac{\partial  H}{\partial z}=0. 
\label{fixed}
\end{equation}
Due to periodicity of the solutions, below we restrict to $\theta\in[0,\,N\pi/2)$. This leads to the following classification:

\begin{itemize} 
\item $\theta={N\pi}/{4}$, and $z$ is a solution of 
\begin{eqnarray*}
 \lambda z + \alpha 
 = \frac{1-3z}{2\sqrt{2(1-z)}} 
%\label{sol1}
\end{eqnarray*}
\noindent which has no solution for $\lambda -\alpha < -1$ while
 there is a unique locally minimal solution for $\lambda -\alpha \geq -1$. 
%In fig. \ref{fig1} we present the graphical solution of this equation for 
% $\theta=\frac{N\pi}{4}$ and different $\lambda - \alpha$ values, illustrating this behaviour
%\vspace{1cm}

%\vspace{1.0cm}
%\begin{figure}[ht]
%\begin{center}
%\epsfig{file=fig1.eps,width=12cm,height=5cm,angle=0}
%\caption{ Graphical solution of the transcendental equation (\ref{sol1}). The crossing between 
%the straight line (left hand side of eq.(\ref{sol1})) and the curve (right hand side of eq.(\ref{sol1})) 
%for $\lambda - \alpha \geq -1$ represents the (unique) solution, while
%for $\lambda - \alpha < -1$ eq.(\ref{sol1}) has no solution. }
%\label{fig1}
%\end{center}
%\end{figure}

\item $\theta=0$, and $z$ is a solution of 

\begin{equation}
 \lambda z + \alpha 
 = \frac{3z-1}{2\sqrt{2(1-z)}} 
\label{sol2}
\end{equation}

\noindent which has a unique locally maximal solution for $\lambda - \alpha < 1$ while for 
$\lambda - \alpha > 1$ there are either two solutions (one locally maximal point and one 
saddle point) or no solutions. 
In Fig. \ref{fig2} we present a graphical solution of (\ref{sol2}).  

\item 
$z=-1$ and $\theta$ is a solution of 
\begin{eqnarray*} 
\cos\left(\frac{4\theta}{N}\right)=\lambda-\alpha
\end{eqnarray*} 
for which there are two saddle point solutions for $|\lambda-\alpha|<1$.  
\end{itemize} 
It is also useful to identify the points $z=1,\,\theta=N\pi/8$ and $z=1,\,\theta=3N\pi/8$, where the 
singular derivative ${\partial H}/{\partial z}$ changes sign. For $z=1$ the Hamiltonian 
(\ref{ham2}) is independent of $\theta$, 
so these points essentially behave like a saddle point.  We remark that (\ref{ham2}) is also independent of $\theta$
for $z=-1$.  
\vspace{1.0cm}

{}From the above we see that there exist fixed point bifurcations for certain choices of the coupling parameters.
These bifurcations allow us to divide the parameter space into four regions, as depicted in Fig. \ref{fig3}.  
The asymptotic form of the boundary between regions I and II is discussed in the Appendix.

\vspace{1.0cm}
\begin{figure}[ht]
\begin{center}
\epsfig{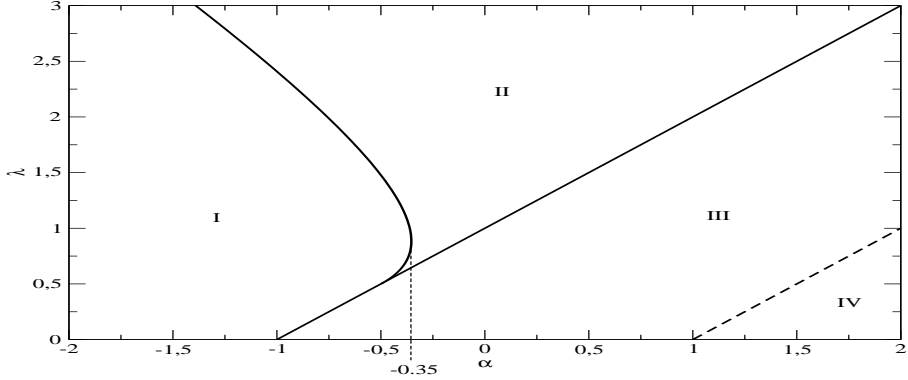}
\caption{Parameter space diagram identifying the different types of solution for  
equation (\ref{fixed}). In region I there are no solutions for $z$ when $\theta = 0$, and one solution for 
$z$ when $\theta = {N\pi}/{4}$. In region II there are two solutions for $z$ when $\theta = 0$, and one solution for 
$z$ when $\theta = {N\pi}/{4}$. In region III exists one solution for $z$ when $\theta = 0$,  one solution for 
$z$ when $\theta = {N\pi}/{4}$, and two solutions for $\theta$ when $z=-1$. In region IV there is one solution for $z$ when $\theta = 0$,  and no solution for $z$ when  
$\theta = {N\pi}/{4}$. The boundary separating regions II and III is given by $\lambda=\alpha+1$, while 
the equation $\lambda =\alpha-1$ separates the regions III and IV.
The boundary between regions I and II has been obtained numerically.}
% For $\theta = \frac{N\pi}{4}$ we change from no solution, region IV, to one solution, others regions, 
% by crossing the straight line 
% ($\lambda - \alpha=-1$). For $\theta = 0$ we have one solution
% in the regions III and IV, two solutions in the region II and no solution in the region I. 
% For $\alpha>\alpha_{c}=-1/\sqrt{8}$ we change from one solution for two solution by crossing the straight line
% ($\lambda - \alpha=1$).  For $-0.5<\alpha<-1/\sqrt{8}$ and $\lambda>0.5$ we change from two solution to no 
% solution and two solution again. For $\alpha<\alpha_{c}=-0.5$ we change from one solution for no solution by 
% crossing the straight line ($\lambda - \alpha=1$) and change from no solution for two solution by crossing the 
% curve of eq.(\ref{lamb1}). }
\label{fig3}
\end{center}
\end{figure} 

To visualise the dynamics, it is useful to plot the level curves of the Hamiltonian (\ref{ham2}).
Since the fixed point bifurcations change the topology of the level curves, qualitative differences can be observed 
between each of the four regions. The results are shown in Figs. ({\ref{level1},\ref{level2}), where for clarity
we now take $4\theta/N\in[-2\pi,\,2\pi]$. 
% , as shown in figs. \ref{level1} and  
% \ref{level2}. 

\begin{figure}[ht]
\begin{center}
\begin{tabular}{cc}
            &             \\
    (a)& (b)    \\
\epsfig{file=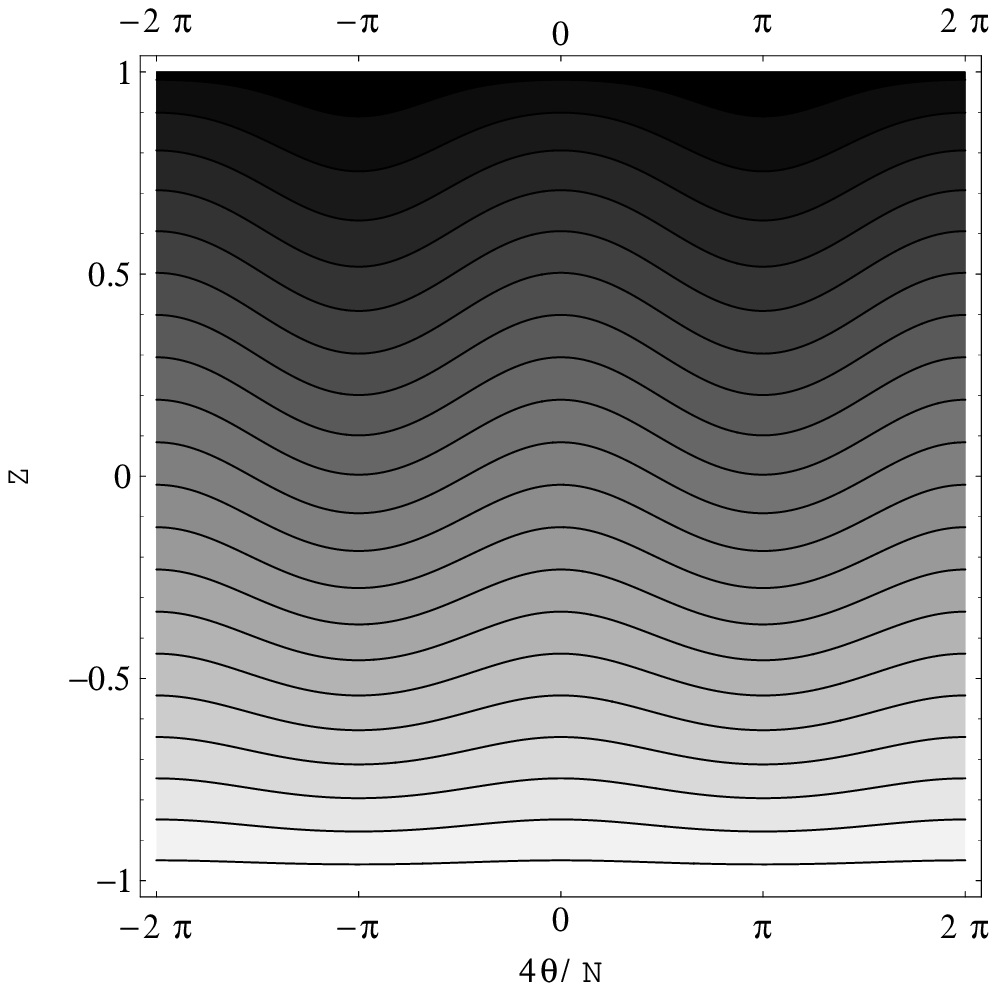,width=6cm,height=6cm,angle=0}&            
\epsfig{file=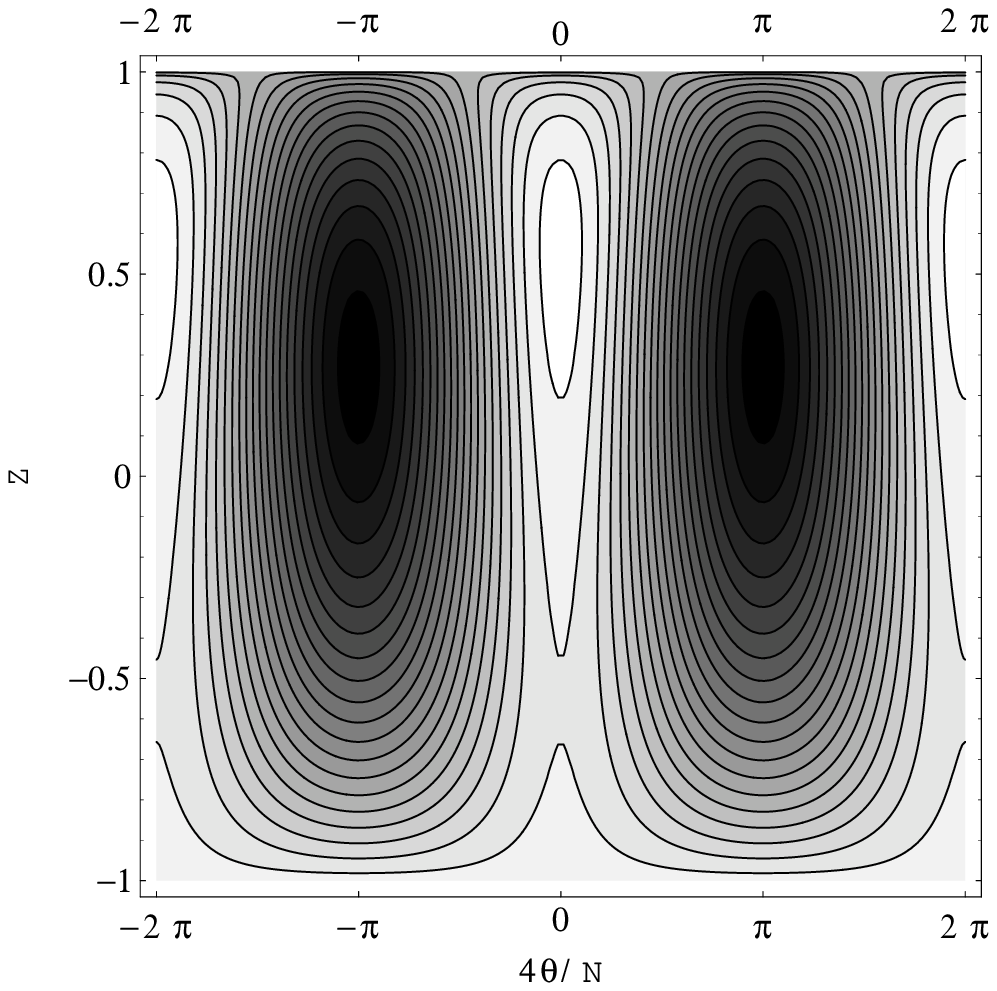,width=6cm,height=6cm,angle=0} \\
\end{tabular}
\end{center}
\caption{Level curves of the Hamiltonian (\ref{ham2}) in (a) region I and (b) region II.
The parameter values are $\lambda=1.0,\,\alpha=-8.0$ for region I 
and $\lambda=1.0,\,\alpha=-0.2$ for region II. 
In region I we observe the presence of local minima for $4\theta/N=\pm\pi$. 
Besides the minima at $4\theta/N=\pm\pi$,
two additional fixed points (a maximum and a saddle point) 
are apparent in region II occurring at $\theta=0$.} 
\label{level1}
\end{figure}

\begin{figure}[ht]
\begin{center}
\begin{tabular}{cc}
            &             \\
    (a)& (b)    \\
\epsfig{file=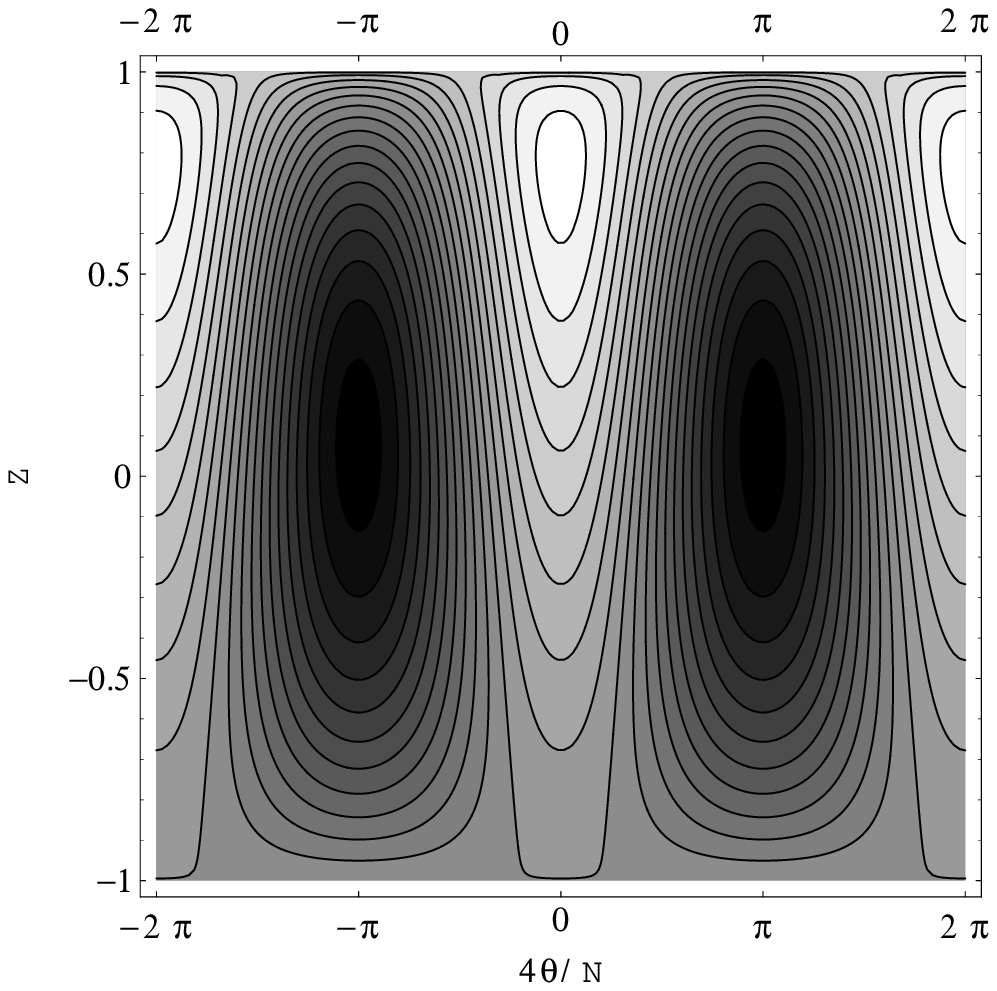,width=6cm,height=6cm,angle=0}&            
\epsfig{file=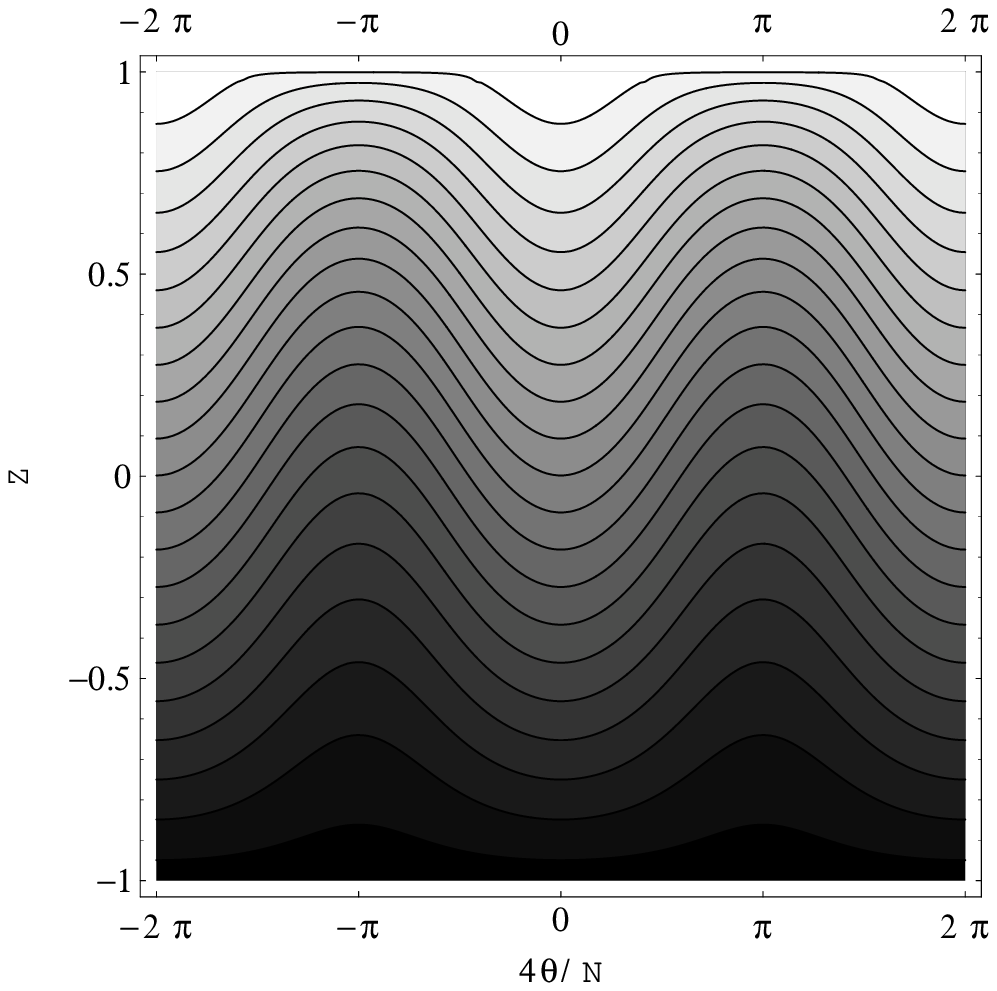,width=6cm,height=6cm,angle=0} \\
\end{tabular}
\end{center}
\caption{Level curves of the Hamiltonian (\ref{ham2}) in  (a) region III and 
(b) region IV. The parameter values are $\lambda=1.0,\,\alpha=0.2$ on the left 
and $\lambda=1.0,\,\alpha=3.0$ on the right. 
In region III we observe the presence of minima at $4\theta/N=\pm\pi$ 
and for $\theta=0$ just one fixed point, a maximum. There are also saddle points for when $z=-1$. 
In region IV just one fixed point (a maximum) occurs for $\theta=0$, which always has $z<1$. In contrast the global 
minimum occurs for $z=-1$.    } 
\label{level2}
\end{figure}

Fig. \ref{level1}(a) shows the typical character of the level curves in region I. The maximal 
level curve occurs along the phase space boundary $z=-1$ and there are two local minima.  
Note that for no 
choice of parameters do these minima occur on the boundary $z=1$, but they may occur arbitrarily close to this boundary. 
If the initial state of the 
system has $z\approx 1$ then $z$ will remain close to 1 for all subsequent times. A similar situation 
is true if the initial state of the system has $z\approx-1$. For both cases we see that the 
evolution of the system is localised.

As the coupling parameters are changed and the system crosses the boundary into region II, two new fixed points,
a maxima and a saddle point, emerge 
at $\theta=0$ which can happen for any $z\in[-1,1)$. On crossing this parameter space boundary the maximum may move towards 
the phase space boundary $z=1$ while the saddle point approaches $z=-1$. Also the two minima may move away from the phase space boundary $z=-1$, as depicted in Fig. \ref{level1}(b). 
The consequence for the dynamics is that for an initial state with $z\approx -1$ the evolution of the system 
is still localised, but for an initial state with $z\approx 1$ the evolution is delocalised. 

Fig. \ref{level2}(a) illustrates what happens when the coupling parameters are tuned to cross over from region II 
into region III. The saddle point at $\theta=0$ approaches $z=-1$, reaching the phase space boundary exactly
when the coupling parameters lie in the boundary between regions II and III.
The saddle point then undergoes a bifurcation into two saddle points occurring at $z=-1$ for different values of 
$\theta$ in region III. The two mimima have also moved away from $z=1$ towards $z=-1$. Now the dynamics is delocalised for both initial states $z\approx 1$ and 
$z\approx -1$.  

It is also possible to tune the parameters to move from region I directly to region III. At the boundary between 
the regions, a single local maximum emerges from the point $z=-1,\,\theta=0$. As the parameters are tuned to move away from the boundary into region III, the maximum moves towards $z=1$ while the minima at $\theta=\pm N\pi/4$ approaches 
$z=-1$.
  
Moving from region III towards region IV causes the two saddle points for $z=-1$ to move towards $\theta=\pm N\pi/4$. 
Again, the two minima for $\theta =\pm N\pi/4$ move towards $z=-1$. Each minimum converges with a saddle point
exactly when the coupling parameters are on the boundary of regions III and IV. Varying the coupling parameters further 
into region IV we find that minima for the Hamiltonian are always at $z=-1,\,\theta=\pm N\pi/4$, and the local maximum
for $\theta=0$ lies close to $z=1$, as shown in Fig. \ref{level2}. For this case the dynamics is localised for both initial states $z\approx 1$ and $z\approx -1$. 

The above discussion gives a general qualitative description of the dynamical behaviour of the classical system in terms 
of the four regions identified in the parameter space. We emphasise that the change in the classical dynamics as the boundary between two regions is crossed is  {\it smooth}.   
Nonetheless, the analysis does give a useful insight into the possible general dynamical behaviours.  
Below we will show that the same holds true for the quantum dynamics.

\section{Quantum dynamics}

Having analysed the classical dynamics, we now want to investigate the extent to which a similar scenario holds for the quantum system. 
For the case $\lambda=0$ (where the coupling for all $S$-wave scattering interactions is zero) the quantum dynamics has previously been studied in \cite{vardi,zlm}. In this instance region II is 
not accessible. It was shown that the dynamics is delocalised for $|\alpha|<1$ and localised otherwise for both atomic and molecular inital states, consistent with 
the classical results described above. A surprising aspect of the classical analysis is the existence of region II where the evolution of a purely molecular inital state is highly localised, whereas the evolution of a purely atomic initial state is completely delocalised. We will see that this also occurs for the quantum case.      
Thus the inclusion of the $S$-wave scattering interactions into the Hamiltonian
gives richer dynamics. 

The time evolution of any state is given 
by $|\Psi(t) \rangle = U(t)|\phi \rangle$, 
where $U(t)$ is the temporal operator $U(t)=\sum_{m=0}^{M}|m\rangle \langle m|\exp(-i E_{m} t)$,  
$|m\rangle$ is an eigenstate with energy $E_{m}$ and $|\phi \rangle =|N_a,N_b \rangle $ represents 
the initial Fock state with $N_a$ atoms and $N_b$ molecules such that $N_a+2N_b=N$.
We adopt the method of directly diagonalising the Hamiltonian as done in \cite{our,ours} for the Bose-Hubbard Hamiltonian (\ref{bh}) and compute the expectation value of the relative number of atoms
$$
%\begin{equation}
\langle N_a(t)-2N_b(t)\rangle=\langle \Psi (t)|N_a-2N_b|\Psi (t)\rangle
%\end{equation}
$$
using two different initial state configurations: a purely atomic state and a purely molecular state.
Hereafter, we will fix the following parameters $N=100, \Omega=1.0, \mu_a =0.0, \mu_b =0.0$ and $U_b=1.0$.

In Fig. \ref{p1} we plot the expectation value of the relative number of atoms 
for $\lambda=1.0$ and 
the choices  $\alpha$ = -8.0, -0.2, 0.2, 3. The  
graphs depict the quantum  dynamics for those cases where the system 
is in regions I, II, III and IV from top to bottom respectively.
On the left we are using a purely atomic initial state $|N,0\rangle $ and on the right hand side a purely 
molecular initial state $|0,N/2\rangle $.

\vspace{1.0cm}
\begin{figure}[ht]
\begin{center}
\epsfig{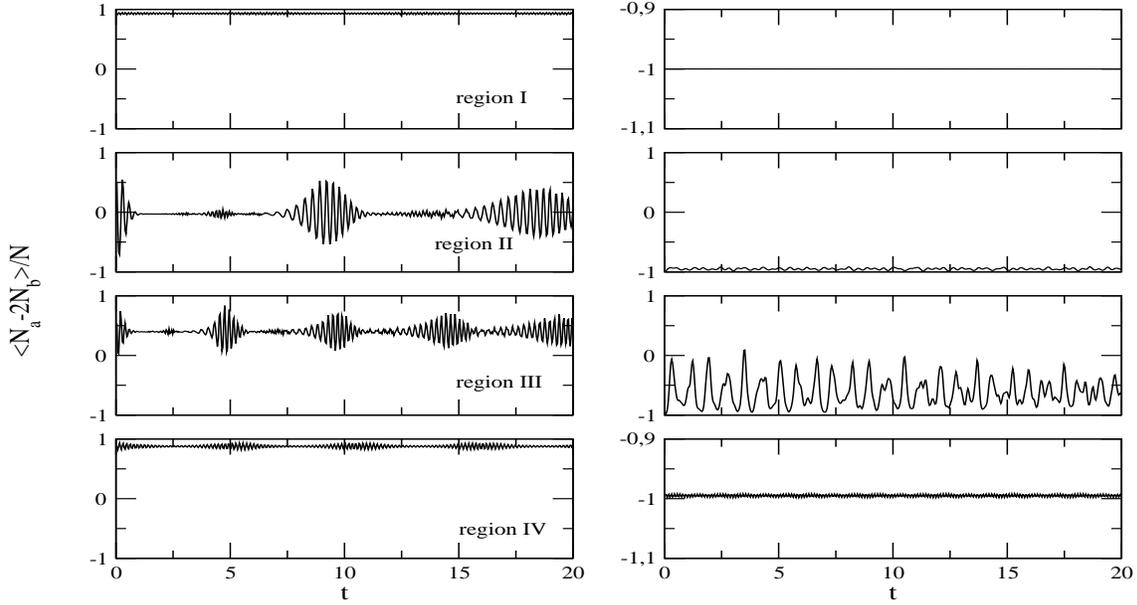}
\caption{Time evolution of the expectation value of the imbalance 
population $\langle N_a-2N_b\rangle/N$ in the four regions defined by the diagram with a purely atomic initial state $|N,0\rangle $ on the left and 
a purely molecular initial state $|0,N/2\rangle $ on the right.
%The central graph in each column represents the dynamics when the system 
%is in a configuration lying exactly at the boundary I-II.
We are using $\lambda=1.0$ and $\alpha=-8.0,-0.2,0.2, 3.0$ (or, in terms of the original variables, $U_b=1, U_{a}=-0.881,0.222,0.278,0.674$ and $U_{ab}=-1.546,0.660,0.774,1.566$).}
\label{p1}
\end{center}
\end{figure}

Figure \ref{p1} displays aspects of the quantum dynamics, such as the collapse and revival of oscillations and 
non-periodic oscillations, which are not features of the corresponding classical dynamics (cf. \cite{mcww} for 
analogous results for the Hamiltonian (\ref{bh})). However it also shows that  
the classification based on classical fixed point bifurcations to determine whether 
the dynamic evolution is localised or delocalised applies to the quantum case. In particular, in region II
it is clear that for an initial atomic state the evolution is completely delocalised, but localised for an initial
molecular state.

%%%%%%%%%%%
\section{Discussion} \label{discussion}

Using the classical Hamiltonian (\ref{ham2}) as an approximation to the quantum Hamiltonian (\ref{ham}),
we have undertaken an analysis to determine 
the fixed points of the system. The bifurcations of the fixed points divide the coupling parameter space into different regions characterising different dynamics, which can also be seen for the quantum dynamics. It is necessary to establish the extent to which the classical approximation is valid.
Since $\lambda$ and $\alpha$ vary with the number of particles,  
it is required that the gap between successive energy levels should approach a 
continuum for large $N$. This imposes that $\lambda,\,\alpha<< N^{3/2}$.     
We can compare this situation to the case of the Bose--Hubbard model (\ref{bh}), where a similar 
classical analysis is valid for $|U/\Omega|<<N$ \cite{leggett}. It was shown in \cite{ours} that for that model
there are transitions in the dynamical behaviour for the quantum regime $|U/\Omega|>>N$, which are not apparent from the classical analysis. These properties were found to be closely related to couplings for when the energy gap between the ground and first excited state was minimal or maximal. We should expect a similar result to occur for (\ref{ham}). 
 
The relationship between fixed point bifurcations and ground-state entanglement has been studied in \cite{hmm05}.
There it was argued that the ground state entanglement of a quantum system will be maximal whenever the classical 
system undergoes a supercritical pitchfork bifurcation for the lowest energy in phase space. A peak in a measure of ground-state 
entanglement has been shown in many instances to be indicative of a quantum phase transition \cite{on,oaff,vidal}. 
For the Hamiltonian (\ref{ham}) 
we have considered here, there are no supercritical pitchfork bifurcations.      
For $\lambda=0$ there is a quantum phase transition at $\alpha=1$, as can be 
seen from the behaviour of certain ground state correlation functions \cite{caok,zlm,hmm03}. This does correspond to a bifurcation of the lowest energy in phase space.  Calculation of the ground-state entanglement in this case have been undertaken in \cite{hmm03} showing that it is maximal at a coupling different from the critical point.    

This is in some contrast to the Bose--Hubbard model (\ref{bh}). 
There, a supercritical pitchfork bifurcation of the lowest energy occurs in the {\it attractive} case
\cite{ks,ours}, and the results of \cite{pd} suggest that indeed the entanglement is maximal at this coupling. (For the repulsive case the ground state entanglement is a smooth, monotonic function of the coupling \cite{hmm03}.) However the 
transition from localisation to delocalisation for the dynamics as studied in \cite{mcww,our,ours} does not occur at the bifurcation. Despite the apparent similarities between (\ref{ham}) and (\ref{bh}), we can see that the inter-relationship between bifurcations of the classical system and properties of the quantum system are very different.

\section*{Acknowledgements}

G.S. and A.F. would like to thank S. R. Dahmen for discussions and CNPq-Conselho Nacional de Desenvolvimento
Cient\'{\i}fico e Tecnol\'ogico for financial support. A.F. also acknowledges
support from PRONEX under contract CNPq 66.2002/1998-99
A.T. thanks FAPERGS-Funda\c{c}\~ao de Amparo \`a Pesquisa do Estado do Rio Grande do Sul for financial 
support. J.L. gratefully acknowledges funding from the Australian Research Council and The University 
of Queensland through a Foundation Research Excellence Award.

%%%%%%%%%%%%%%
\section*{Appendix}

In this Appendix we analyse the boundary dividing regions I and II.
In particular, we determine the asymptotic relation between $\lambda$ and $\alpha$ 
when $\lambda$ is large and when $\lambda$ is close to 1/2. 
We also compute the maximum value of $\alpha$ on this boundary.

Consider
\begin{eqnarray*}
f(z)&=&\lambda z + \alpha \\
g(z)&=&\frac{3z-1}{2\sqrt{2(1-z)}}
\end{eqnarray*}
where the fixed points occur when $f(z)=g(z)$. 
We want to determine the boundary between the cases when there is no solution 
and two solutions.
This boundary is given by the case when $f(z)$ is the tangent line to $g(z)$. Now

$$
\frac{dg}{dz} = \frac{1}{2\sqrt{2}}(3(1-z)^{-1/2}+\frac{1}{2}(3z-1)(1-z)^{-3/2})
$$
so $z$ is determined by the condition
\begin{equation}
\lambda = \frac{dg}{dz}. \label{alsouseful}
\end{equation}

Below we consider three cases: 
\begin{itemize}
\item[(i)] First put $z=-1+u$ where $u$ is small and positive. Then
$$
\frac{dg}{dz}%& =& \frac{1}{2\sqrt{2}}(3(2-u)^{-1/2} + \frac{1}{2}(-4+3u)(2-u)^{-3/2}) \\
%& \approx & \frac{1}{2\sqrt{2}}(\frac{3}{\sqrt{2}}(1+\frac{1}{4}u) - \frac{1}{\sqrt{2}}(1-\frac{3}{4}u)(1+\frac{3}{4}u)) \\
 \sim  \frac{1}{2} + \frac{3}{16}u. 
$$

Solving for $u$ gives
$$
u \sim \frac{8}{3}(2\lambda - 1).
$$
Now we need

\begin{eqnarray*}
f(z) &=& g(z)  \\
\lambda(-1+u) + \alpha 
     & = & \frac{1}{2\sqrt{2}}(-4+3u)(2-u)^{-1/2}  \\
     & \sim & -1 + \frac{1}{2}u.
\end{eqnarray*}

We can substitute in $u$ to find a relation between $\lambda$ and $\alpha$:

$$
\alpha  \sim  -\frac{1}{2} +\left(\lambda-\frac{1}{2}\right) -\frac{16}{3}\left(\lambda-\frac{1}{2}\right)^2 
$$

This curve is valid for $(\lambda -1/2)$ positive. Also

$$
\left. \frac{d\alpha}{d\lambda} \right|_{\lambda=1/2}  =  1
$$
so the curve separating regions II and III is tangential to the curve separating regions I and III
at $\lambda=1/2$. 

\item[(ii)] Next we look at the case when $z=1 - u$ with $u$ small and positive. Here we find

\begin{eqnarray*}
g & = & \frac{2-3u}{2\sqrt{2u}}  \\
     & \sim & \frac{1}{\sqrt{2u}},  \\
\frac{dg}{dz} &\sim&  \frac{1}{2\sqrt{2}u^{3/2}}
\end{eqnarray*}
so that
$$
u \sim \frac{1}{2}\lambda^{-2/3}.
$$
This leads to 
\begin{eqnarray}
\alpha & = & g(z) - \lambda z \label{useful} \nonumber \\
%       & \sim & \frac{1}{\sqrt{2u}} - \lambda(1-u)  \nonumber \\
       & \sim & -\lambda + \frac{3}{2}\lambda^{1/3} \label{ass2}.
\end{eqnarray}
The asymptotic equation (\ref{ass2})
is valid for large positive  values of $\lambda$.

\item[(iii)] To complete the picture, finally we investigate the maximum of $\alpha$ with respect to $\lambda$. 
{}From (\ref{alsouseful},\ref{useful}) we have 
\bea
\frac{d\alpha}{d\lambda} & = & \frac{dg}{dz}\frac{dz}{d\lambda} - \lambda\frac{dz}{d\lambda} - z  \\
                         & = & -z
\eea

so the maximum occurs at $z=0$. Looking at the asymptotic behaviour around $z=0$
we have

\begin{eqnarray*}
g(z) %& \approx & -\frac{1}{2\sqrt{2}}(1-3z)(1 + \frac{1}{2}z + \frac{3}{8}z^{2})  \\
     &\sim& -\frac{1}{2\sqrt{2}}(1 - \frac{5}{2}z - \frac{9}{8}z^{2})  \\
\frac{dg}{dz} &\sim&  -\frac{5}{4\sqrt{2}}(1 + \frac{9}{10}z)
\end{eqnarray*}
which gives
$$
z  \sim  \frac{10}{9}(\frac{4\sqrt{2}}{5}\lambda -1).
$$
Using this we can find an expression for $\alpha$ in terms of $\lambda$:

$$
\alpha %& \approx & -\frac{1}{2\sqrt{2}} - \frac{9}{16\sqrt{2}}z^{2}  \\
\sim -\frac{1}{2\sqrt{2}} - \frac{25}{36\sqrt{2}}(\frac{4\sqrt{2}}{5}\lambda - 1)^{2}\label{ass3}
$$ 
The first term above corresponds to the maximal value of $\alpha\approx -0.35$
as depicted in Fig. \ref{fig3}. 
\end{itemize}

\end{document}